\documentclass[10pt,aps,prd,nofootinbib,superscriptaddress, twocolumn,preprintnumbers,balancelastpage]{revtex4}
\usepackage{amssymb,amsmath,latexsym,graphics, graphicx,epsfig,multirow,comment,hyperref,appendix}

\def\EE{{\cal E}}

\newcommand{\GeV}{\,\mathrm{GeV}}

\newcommand{\keV}{\,\mathrm{keV}}

\newcommand{\half}{{\frac{1}{2}  }}

\newcommand{\esc}{{\text{esc}}}

\newcommand{\vesc}{{v_{\text{esc}}}}

\newcommand{\gsim}{\lower.7ex\hbox{$\;\stackrel{\textstyle>}{\sim}\;$}}
\newcommand{\lsim}{\lower.7ex\hbox{$\;\stackrel{\textstyle<}{\sim}\;$}}
\newcommand{\kms}{\text{ km s${}^{-1}$}}

\usepackage[usenames,dvipsnames]{color}
\definecolor{orange}{RGB}{220,120,0}

\def\apjs{ApJ\&S}

\begin{document}

\title{The Dark Matter at the End of the Galaxy
 }

\author{Mariangela Lisanti}
\affiliation{PCTS, Princeton University, Princeton, NJ 08540}
\affiliation{Particle \& Particle Astrophysics Department, SLAC National Accelerator Laboratory, Menlo Park, CA 94025}

\author{Louis E. Strigari}
\affiliation{Kavli Institute for Particle Astrophysics and Cosmology \& Physics Department, Stanford University, Stanford, CA 94305}

\author{Jay G. Wacker}
\affiliation{Particle \& Particle Astrophysics Department, SLAC National Accelerator Laboratory, Menlo Park, CA 94025}

\author{Risa H. Wechsler}
\affiliation{Particle \& Particle Astrophysics Department, SLAC
  National Accelerator Laboratory, Menlo Park, CA 94025}
\affiliation{Kavli Institute for Particle Astrophysics and Cosmology \& Physics Department, Stanford University, Stanford, CA 94305}

\begin{abstract}
  Dark matter density profiles based upon $\Lambda$CDM cosmology motivate an
  {\it ansatz} velocity distribution function with fewer high velocity
  particles than the Maxwell-Boltzmann distribution or proposed variants.  
  The high velocity tail of the distribution is  determined by the outer slope of the dark matter halo 
  -- the large radius behavior of the Galactic dark matter density. 
N-body simulations of Galactic halos reproduce the high velocity behavior of this {\it ansatz}. 
Predictions for direct detection rates are dramatically affected for models where the threshold 
scattering velocity is within 30\% of the escape velocity.
\end{abstract}
\preprint{SLAC-PUB-14273}
 \maketitle

\section{Introduction}

Dark matter is the dominant form of matter in the Universe and
measuring its properties is one of the most important outstanding
problems in astrophysics and particle physics.  Directly detecting
Galactic dark matter through terrestrial experiments is probably the
least ambiguous method for determining its identity.  Accurately
predicting scattering rates for direct detection experiments requires
a model for the velocity distribution of Galactic dark matter.  This
article proposes a class of velocity distribution functions that are
motivated by cosmological models for the Galactic dark matter and
studies how this {\it ansatz} affects predictions for direct detection
experiments.

Dark matter halos form through a continuous process of smooth accretion
and merging of smaller mass halos in the $\Lambda$CDM model of
cosmology.  Clues about the formation history of these halos can be
found in their phase space distributions, motivating a careful study
of dark matter densities and velocities.  Modern numerical
simulations, which probe a large range of halo mass scales, have shown
that the merging process leads to a near universal double-power-law
density profile that is approximately described by the
Navarro-Frenk-White (NFW) model~\cite{nfw}. Though
the physical origin for NFW-like profiles is unknown, 
they provide a valuable link between the small scale properties of dark
matter halos and the cosmological model on large scales.

The velocity distribution is dynamically related to the density
profile for equilibrium dark matter halos~\cite{Eddington,BT08}.
In direct dark matter detection studies, the `Standard Halo Model'
(SHM) \cite{Drukier:1986tm} is the canonical velocity distribution for
interpreting results from experiments and constraining dark matter
models~\cite{many}.  Isothermal and isotropic approximations to the
dark matter distribution functions such as the SHM are useful because
they simplify rate calculations.  However, recent high resolution
cosmological simulations provide strong evidence that the SHM does not
appropriately capture the behavior of the dark matter in the Galactic
halo
\cite{Vogelsberger:2008qb,Kuhlen:2009vh,Fairbairn:2008gz,Ling:2009eh}.
By mapping out the dark matter phase space distribution, these N-body
simulations consistently point to deviations from a standard
Maxwellian velocity distribution function, most prominently at the
tail of the distribution~\cite{LeonardTremaine}.  This new evidence
motivates a reevaluation of the SHM and careful consideration of
velocity distributions that better capture the features of the
Galactic halo.

There have been recent attempts to derive the dark matter velocity
distribution from first principles~\cite{Hjorth:2010mn} or from
observations of rotation curves~\cite{Chaudhury:2010hj}.  However, the
resulting models over-predict the number of high velocity particles
compared to the highest resolution dark matter-only numerical
simulations with $\sim 10^3$ M$_{\odot}$ particles
\cite{Vogelsberger:2008qb, Kuhlen:2009vh}, as well as simulations that
include baryons~\cite{Ling:2009eh}.  It has been claimed that Tsallis
distributions, derived from non-extensive statistical
mechanics~\cite{Hansen:2004dg}, provide a good fit for the baryon plus
dark matter simulation in~\cite{Ling:2009eh}.  However, because the
resolution for this simulation is $\sim 7\times10^5$ M$_{\odot}$, it is 
limited at probing the high velocity tail relative to the dark matter only simulations.  An
alternate approach is to fit the numerical results with an arbitrary
function~\cite{Kuhlen:2009vh, Fairbairn:2008gz}; while these models
are useful for making experimental predictions, there is no clear motivation
for the fit parameters.

This paper obtains a velocity distribution function that describes the double power-law density models.  While these velocity
distributions have been obtained numerically for isotropic models and
also for certain classes of spherical anisotropic
models~\cite{Widrow2000,Evans:2005tn,solutions}, no
analytic solution has been found.  The isotropic velocity distribution
function corresponding to double power-law density profiles~\cite{nfw}
is described by
\begin{equation}
f_k(v) \propto \left[ \exp\left( \frac{ v_\text{esc}^2 -v^2}{ {k}\, v_0^2}\right) - 1\right]^{{k}}  \Theta(\vesc - v),
\label{eq:vdist}
\end{equation}
where ${k}$ is the power-law index, $\vesc$ is the escape velocity,
and $v_0$ is the dispersion. This distribution function is significantly different from Maxwell-Boltzmann distributions and its variants in the high velocity tail.  It describes only the smooth dark matter
component; modifications to the distribution function from discrete
subhalos and streams can be important phenomenologically but need to
be added  separately~\cite{Helmi:2002ss,Stiff:2003tx,Kamionkowski:2008vw}.  Additionally, deviations from a smooth distribution arise from gravitational scattering from the Sun and the Jovian planets~\cite{Peter:2009mm}.

Direct dark matter detection experiments look for low energy nuclear recoil events
caused by dark matter scattering off of target nuclei~\cite{Lewin:1995rx}.  The nuclear
recoil thresholds for dark matter detection experiments are in the
range of tens of $\keV$,
which implies that the recoil energies in many models of dark matter are well below threshold. 
Only particles with the largest velocities have sufficiently energetic nuclear recoils to
be visible at direct detection experiments. This fact makes it
particularly important to accurately describe the high velocity tail
of the dark matter velocity distribution in the halo
surrounding the Milky Way.  Relative to a Maxwell distribution, (\ref{eq:vdist}) predicts a smaller fraction of dark matter near the
escape velocity, which affects the relative rates between direct detection experiments, 
especially those with different target nuclei or energy thresholds.

This paper is organized as follows.   Sec.~\ref{sec:theory} motivates 
the {\it ansatz} in \eqref{eq:vdist} and derives a
general form for the power-law index.  Sec.~\ref{sec:baryons}
compares \eqref{eq:vdist} to numerical calculations of the full
velocity distribution function for Milky Way-like spatial profiles,
including a gravitational potential model of the disk.  The ansatz in \eqref{eq:vdist} is also compared to the results of N-body simulations.  Elementary
direct detection phenomenology is explored in Sec.~\ref{sec: directdetection}, and conclusions are presented in
Sec.~\ref{sec:outlook}.

\section{Equilibrium Dark Matter Distributions}
\label{sec:theory}

The Galactic halo forms through a
process of hierarchical merging, smooth accretion, and violent relaxation~\cite{nfw,Wechsler:2001cs,Wang:2008un}. 
The central regions of halos, where relaxation processes have ended, are in quasi-static equilibrium~\cite{LyndenBell:1966bi}. 
In this case the dominant component of Galactic dark matter is
described by a steady-state distribution function that evolves slowly
with time.  
In the outer regions of the halo, at low binding energies,
the phase space
distribution will depend more dramatically on both the history and
environment of the halo.  For example, a halo with a fairly quiescent
formation history will have a different phase space distribution at
low binding energies compared to one with a recent major merger
\cite{Vogelsberger:2008qb}.
Sec.~\ref{sec:baryons} argues that the behavior of the distribution function of the Milky Way at high velocities and low binding energies resembles an equilibrated system on average. 
There can be large excursions from equilibrium at the high velocity tail due to recent
accretion; however, N-body simulations suggest that these are spatially localized streams.  

According to Jeans theorem, the six-dimensional phase space distribution function $f(\vec{x}, \vec{v})$ for a spherically symmetric and isotropic system can be written in terms of the energy, which is an integral of motion:
\begin{eqnarray}
\EE \propto 2 \psi(\vec{x}) -  \vec{v}^2. 
\end{eqnarray}
The local escape velocity is defined as the velocity  where $\EE=0$
and therefore
\begin{eqnarray}
\EE \propto  v_{\esc}^2(r) - v^2 .
\label{eq: velocity}
\end{eqnarray}
Due to the assumption of equilibrium, $f(\EE)$ for $\EE <0$ must vanish since this corresponds
to particles with velocities greater than the escape velocity of the system.

If the dark matter of the  Milky Way is nearly equilibrated, then the Jeans theorem implies that
the dark matter density determines the velocity distribution function.
Dark matter halos in cosmological simulations are well fit by spherically
averaged double power-law density distributions of the form, 
\begin{eqnarray}
\rho(r) = \frac{\rho_s}{(r/r_s)^\alpha (1 + (r/r_s))^{(\gamma-\alpha)}},
\label{Eq:DoublePower}
\end{eqnarray}
where $\rho_s$ is the scale density, $r_s$ is the scale radius,
$\alpha$ is the slope of the halo density near the core, and $\gamma$
is the slope at large radii.  For example the NFW profile has
$(\alpha, \gamma) = (1,3)$~\cite{nfw}, while the Hernquist model has
$(\alpha,\gamma)= (1,4)$~\cite{Hernquist:1990be}.  The Jaffe model has
$(\alpha,\gamma)=(2,4)$, but the $\alpha=2$ inner slope is steeper
than indicated by numerical simulations.  

Velocity distribution
functions that behave as Maxwell-Boltzmann distribution functions at
low velocities 
arise from dark matter halos with inner slopes of $\alpha =2$.
This article is primarily concerned with the effects of the
dark matter velocity distribution on direct detection rates.  The
relative velocity between dark matter and target nuclei for $v \ll
v_0$ is set by the solar velocity; therefore, the low velocity
behavior of the Galactic dark matter does not significantly affect
direct detection rates.  Though finding a velocity distribution
function that describes $\alpha < 2$ is important, these modifications
to the dark matter velocity do not qualitatively change the direct
detection rates \cite{Kamionkowski:1997xg}.  The goal of this section
is to derive a phenomenological velocity distribution function that
can arise from these double power-law density models to predict direct
dark matter detection rates.


A given density distribution is related to the velocity distribution through the gravitational potential of the halo, 
$\psi(r)$:
\begin{equation}
\nabla^2 \psi = - 4 \pi G \rho = - 4 \pi G M \int f(v) d^3v,
\label{eq:poisson}
\end{equation}
where $G$ is the gravitational constant and $M$ is the halo mass.  For an isotropic density distribution, 
there is a one-to-one correspondence between the density and ergodic distribution function, given by the Eddington formula:
\begin{eqnarray}
f(\psi) = \frac{1}{\sqrt{8} \pi^2} \Bigg[ \int^{\psi}_0  \frac{ d\psi'}{\sqrt{ \psi - \psi'}}  \frac{ d^2 \rho}{d\psi'{}^2} +  \frac{1}{\sqrt{\psi}} \Big(\frac{d\rho}{d\psi}\Big)_{\psi = 0} \Bigg].
\label{eq:Eddington}
\end{eqnarray}
The expression for $f(\psi)$ can be written in terms of the binding
energy $\EE$ by replacing $\psi\rightarrow \EE$.  Equation
(\ref{eq:Eddington}) is a powerful simplification that allows one to
solve for the distribution function of an arbitrary spherical density
model.  In many cases, however, the solution for $f(v)$ is
analytically intractable and must be obtained by solving
(\ref{eq:Eddington}) numerically.  This is the case for most
combinations of $(\alpha, \gamma)$ in (\ref{Eq:DoublePower}); as a
result, there is no general closed-form analytic solution for the
velocity distribution function that corresponds to double power-law
density models.  Fortunately, it is still possible to assume an {\it
  ansatz} for $f(v)$ that reproduces the numerical solution of the
Eddington formula for double power-law densities.  The possibilities
for such an {\it ansatz} are limited if one assumes that the dark
matter halo is in equilibrium.

The high velocity tail of the local dark matter distribution function arising from equilibrated, double power-law models is in conflict with the Standard Halo Model and its variants.
Consider the following generalization of the Standard Halo Model as an {\it ansatz}
\begin{equation}
f(\EE) \propto (e^{\EE/\EE_0} - 1)^{{k}} \Theta(\EE), 
\label{eq:ansatz}
\end{equation}
where ${k}$ is the power-law index of the distribution and describes the behavior near the escape velocity.  
The benefit of this {\it ansatz} is that it 
satisfies the Jeans theorem for an equilibrated system and goes continuously to zero at the escape velocity.  

The power-law index is defined as
\begin{equation}
k = \lim_{\EE\rightarrow 0^+}  {k}(\EE),
\end{equation}
with
\begin{equation}
{k}(\EE) \equiv \frac{ \EE}{ f(\EE)} \frac{ d f(\EE)}{d \EE}.
 \end{equation}
 For double-power density profiles of the form in
 \eqref{Eq:DoublePower}, the power-law index can be evaluated
 analytically.  In this case, the second term of the Eddington formula
 is negligible. Expanding the density, potential, and the distribution
 function in \eqref{eq:Eddington} around small $\cal{E}$ gives a power-law index
\begin{equation}
{k} = \gamma - \frac{3}{2}
\label{eq:eta}
\end{equation}
for $\gamma >3$ \cite{Kochanek:1995xv}.  As $\gamma \rightarrow 3$,
${k}(\EE)$ does not approach a simple power-law and the approximations
that lead to \eqref{eq:eta} break down~\cite{Widrow2000}.  
Sec.~\ref{sec:baryons} fits the numerical solutions to the parameters for the {\it ansatz} velocity distribution function in \eqref{eq:vdist}.  As
$\gamma \rightarrow 3$, the best fits for ${k}$ tend to be $k\simeq
2.0$, slightly larger than \eqref{eq:eta}. Earlier studies from galaxy
formation via violent relaxation motivated
${k}=1.5$~\cite{Tremaine:1987}.

It is remarkable that $k$ takes such a general and simple form.  The
index is determined almost exclusively by the outer slope of the
density distribution; all terms that depend on $\alpha$ vanish in the
low energy limit.  The outer slope controls the behavior of
${k}$ because the dark matter particles with the highest velocities
are those with the smallest binding energies.  These particles
will be in highly energetic orbits about the halo, and will be
concentrated at large radii, far from the core.  Density distributions
with larger outer slopes have fewer particles orbiting at large radii,
which means that the low-energy component of $f(\EE)$ is suppressed.
A large value of $k$ precisely captures this behavior.

The distribution function for binding energies can be rewritten in terms
of velocities using the relation in \eqref{eq: velocity} to get
\eqref{eq:vdist}.  The velocity distribution {\it ansatz} in
\eqref{eq:vdist} is well-described by a Gaussian peaked near $v_0$ for
$v \ll v_{\text{esc}}$.  As $v\rightarrow v_{\text{esc}}$, the
distribution function approaches
\begin{eqnarray}
 f(v)  \rightarrow (v_{\text{esc}} - v)^{{k}}. 
 \label{eq:fv_approx}
\end{eqnarray}
Cosmological N-body simulations indicate 
$\gamma \sim 3 -5$~\cite{nfw,Busha:2004uk}, which means that the velocity distribution
falls off near the escape velocity to the power ${k} = [1.5, 3.5]$.  A
more detailed comparison with cosmological simulations follows in
Sec.~\ref{sec:baryons}.

The Standard Halo Model and the King model provide a useful comparison to the distribution in \eqref{eq:ansatz} and they
are defined, respectively, as
\begin{eqnarray}
\nonumber
f_{\text{SHM}}(\EE) &=&  N(\EE_0) e^{ \EE/\EE_0}\Theta(\EE)\\
f_{\text{King}}(\EE) &=& N(\EE_0) (e^{\EE/\EE_0} -1)\Theta(\EE). 
\label{eq:SHM_King}
\end{eqnarray}
These distributions are frequently used for direct dark
matter detection predictions because they make the calculations tractable. 
In addition, they satisfy Jeans theorem under the assumption of isotropy
and spherical symmetry.  However, they do not correspond to NFW-like
density models, especially near the high-velocity tails of the
distributions.  In particular, the SHM behaves near the tail as ${k}
\rightarrow 0$, and the King model has ${k} =1$.  Consequently, these
velocity distribution functions over-predict the number of particles
in the tail of the distribution.

The Tsallis distribution is another model for the velocity distribution that has been recently discussed in the literature, and is
defined as      
 \begin{equation}
 f_{\text{Tsallis}}(v) \propto \Bigg(1-(1-q) \frac{v^2}{v_0^2}\Bigg)^{q/(1-q)}. 
 \end{equation} 
 The Tsallis distribution predicts that the escape velocity is given by $v_{\text{esc}}^2 = v_0^2/(1-q)$ and
 $k= q/(1-q)$, where the distribution in \eqref{eq:vdist} sets these three parameters independently.  
 Hansen et al.~\cite{Hansen:2005yj} show there is a
 correlation between the parameter $q$ and the local density slope,
 implying that $q$ varies with radius if the density slope is not the
 same at all radii. A disadvantage of this model, however, is that does not satisfy the Jeans theorem for spherical
 and isotropic systems if the circular velocity, $v_0$, is held
 constant.  This violation of Jeans theorem is also true for the
 generalized Maxwellian distribution, which has been used to model the
 radial and tangential components of the velocity
 distribution~\cite{Fairbairn:2008gz}.
 
 The formalism and models in this section  only apply to 
 the spherically-averaged velocity distribution, and do not capture any physics
 pertaining to steams, subhalos, or any other structure in the phase
 space distribution. Though streams are unlikely to affect the overall
 velocity distribution because their densities are less than $\sim
 0.1\%$ of the smooth dark matter density~\cite{Vogelsberger:2008qb},
 they are seen to occupy the high energy tail of the velocity
 distribution~\cite{Helmi:2002ss,Maciejewski:2010gz}.

\section{Numerical Evaluation} 
\label{sec:baryons}

The previous section derived how the high velocity tail of the dark
matter phase space is populated for cosmologically-motivated spatial
densities.  The {\it ansatz} in \eqref{eq:vdist} interpolates between
a Maxwellian distribution at low velocities to one matching the high
velocity behavior in \eqref{eq:eta}.  This section compares the
analytic approximation in \eqref{eq:eta} with the velocity
distributions that are calculated from the full numerical solution of
the Eddington formula in \eqref{eq:Eddington}.  Additionally, it shows
that the {\it ansatz} distribution function accurately describes the
velocity distributions of simulated dark matter halos both with and
without baryonic physics.

\subsection{Distribution Function Model for the Milky Way} 

A full gravitational potential model for the Milky Way 
is necessary to evaluate the Eddington formula, which requires including a model for the baryons.
The potential model is comprised of the sum of three components: 
the dark matter, the disk, and the bulge, 
so that $\psi_{\text{tot}} = \psi_{\text{halo}} + \psi_{\text{disk}} + \psi_{\text{bulge}}$.
The halo mass within a given radius is obtained from \eqref{Eq:DoublePower}, 
while the velocity of the local standard of rest (LSR) and the escape velocity
are given by
\begin{eqnarray}
v_{\text{lsr}}^2 =  - r \frac{d\psi_{\text{tot}}}{dr} 
 \quad \quad
\vesc = \sqrt{2 | \psi_{\text{tot}}|},
\end{eqnarray}
respectively.

\begin{figure}[t] 
   \centering
   \includegraphics[width=3.5in]{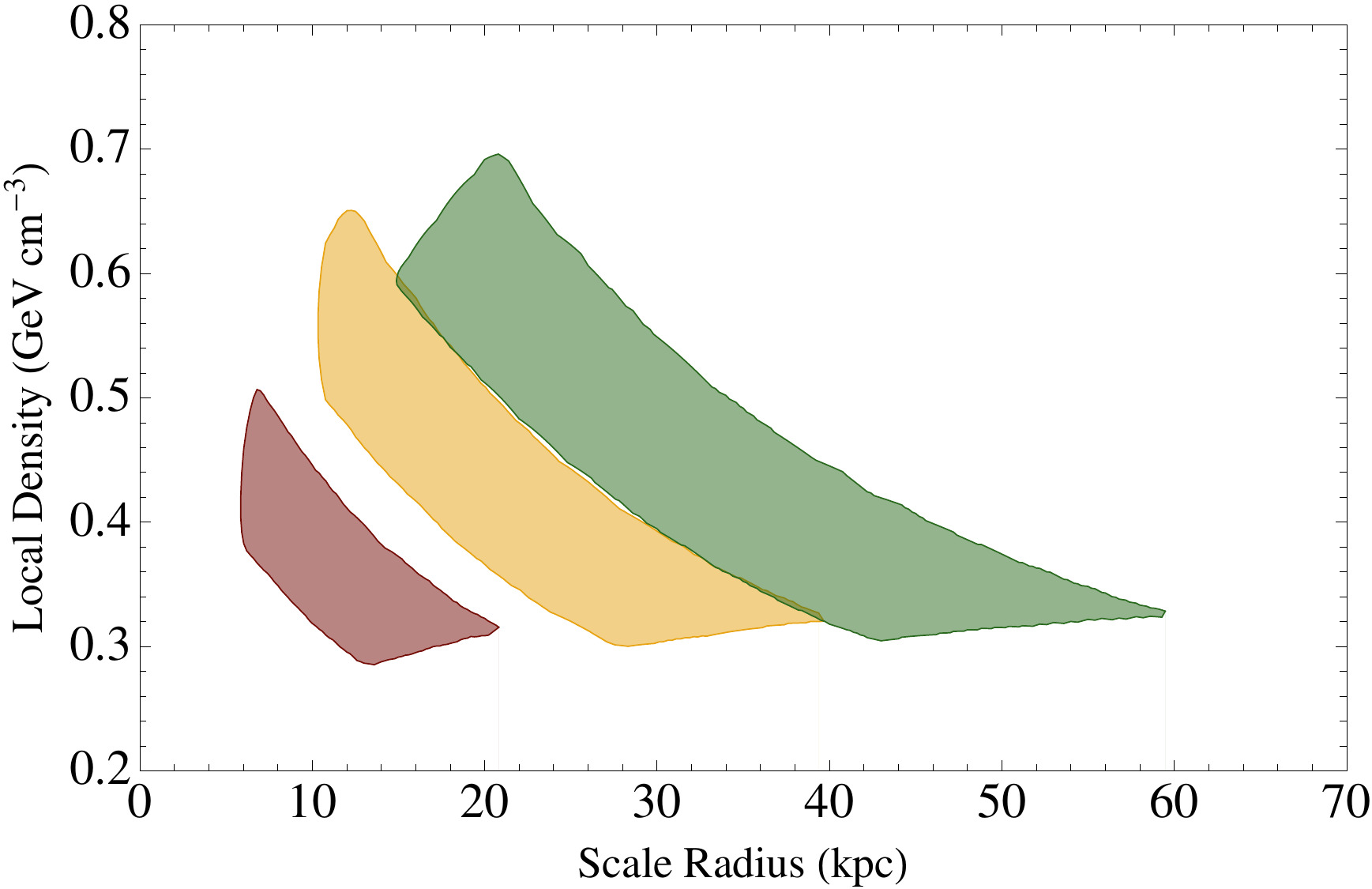} 
   \caption{Region of parameter space in local density (defined at 8.5
     kpc) and scale radius, $r_s$, consistent with (i) halo mass internal to
     60 kpc in the range $(4\pm 0.7)\times10^{11}$ M$_\odot$, (ii) escape
     velocity in the range 450-600 km s$^{-1}$, and (iii) circular velocity
     in the range 200-280 km s$^{-1}$, for models with $\alpha = 1$ and
     $\gamma = 3$ (red), $4$ (yellow), and $5$ (green).}
   \label{fig:regions}
\end{figure}

The value of $\psi_{\text{halo}}$ is uniquely determined for a given pair, ($\alpha$,$\gamma$), 
and for specific values of the scale radius, $r_s$, and scale density, $\rho_s$. 
The velocity distribution {\it ansatz} in~\eqref{eq:vdist} is compatible with 
any value of the outer slope with  $\gamma >3$,
while the inner slope is given by $\alpha=2$.
However, the behavior of the potential for $r \ll r_s$ dominantly determines the low velocity behavior
of Galactic dark matter and does not significantly affect predictions for direct detection rates. 

The full model for $\psi_{\rm total}$ needs to include additional
potentials for the bulge and disk components.  Here, a spherically-symmetric
potential is used for the bulge component of the form
\begin{eqnarray}
\psi_{\rm bulge} = -GM_{\rm bulge}/r,
\end{eqnarray} 
where the mass of the bulge is $M_{\rm bulge} = 10^{10} \, M_\odot$ \cite{BT08}. 
The disk is modeled by a spherical potential of the form
\begin{equation}
\psi_{\rm disk} = -GM_{\rm disk}(1-e^{-r/r_d})/r.
\label{eq:disk}
\end{equation}
The potential formula in \eqref{eq:disk} contributes  $\sim 115 \kms$ to the circular velocity for a disk consisting of  gas and stars with  $M_{\rm disk} = 5 \times 10^{10}$ M$_\odot$ and a scale length of $r_d = 5$ kpc. 
The spherical approximation in~ \eqref{eq:disk} is the monopole term for a standard axisymmetric 
double exponential disk potential \cite{BT08}.  
In comparison, for the assumed disk mass, 
the axisymmetric double exponential disk potential gives $v_{\text{lsr}} \simeq 130\kms$ at the solar radius, 
and is 11\% larger than $v_{\text{lsr}}$
derived from the spherical approximation.  
The disk potential sets the circular velocity of the Milky Way that arises from non-dark matter, {\it i.e.} visible matter. 

The measured values of $v_{\rm lsr}$, $v_{\rm esc}$, and the
integrated halo mass provide constraints on the parameters of the
velocity distribution function.  
The circular velocity is in the range~\cite{v0}
\begin{eqnarray}
v_{\rm lsr} = [200, 280]\kms .  
\end{eqnarray}
The 90\% confidence level on the escape velocity is in the range 
\begin{eqnarray}
v_{\text{esc}} = [498, 608]\kms, 
\end{eqnarray}
 and is determined from local high velocity stars by the RAVE survey~\cite{Smith:2006ym}. 
This constraint on the escape velocity is only attainable for a prior on
the spectral index for the velocity distribution of halo stars, 
\begin{eqnarray}
f_\star(v) \propto (v_{\rm esc} - v)^{k_\star},
\end{eqnarray}
 with 
\begin{eqnarray}
\label{eq:kstars}
k_{\star} = [2.7, 4.7].
\end{eqnarray}
Simulations indicate that the power-law index for halo stars is
steeper than dark matter.  The values of $k_{\star}$ in
\eqref{eq:kstars} correspond to ${k} = [0.5, 2.5]$ for dark matter
\cite{Smith:2006ym}.  The lower values of $k$ correspond to unphysical
halos, with $k_\star$ near the upper end of the range in
\eqref{eq:kstars}. Since the escape velocity is correlated 
with the spectral index of the stars and the dark matter, 
if values of $k$ and $k_\star$ on the lower end of the allowed
region are used, lower values of $v_{\text{esc}}$ must be used, {\it i.e.}
$v_{\text{esc}} \simeq [450, 550] \kms$.

Finally, there is a bound on the integrated dark matter mass of the
Milky Way halo within 60 kpc, determined by distant halo stars to be~\cite{Xue:2008se}
\begin{eqnarray} 
M(< 60 \, {\rm kpc}) = (4.0 \pm 0.7) \times 10^{11} M_\odot .
\end{eqnarray}
In the $\Lambda$CDM model, an extrapolation of the mass profile to larger
radii implies a total Milky
Way halo mass of $\sim 10^{12}$ M$_\odot$~\cite{Xue:2008se,Li:2007eg}. 
The above constraints are
examined in the context of the power-law parameterization in
\eqref{Eq:DoublePower}; more generic halo profiles are beyond the
scope of this analysis, see~\cite{mcmc} for further discussion.

Fig.~\ref{fig:regions} shows the regime of scale radius-local dark matter density 
parameter space that is consistent with the observational bounds above, where the local dark matter
density is simply defined  as $\rho(r=8.5 \, {\rm kpc})$ in \eqref{Eq:DoublePower}.
Three different
halo models are shown with different values of the outer slope of $\gamma = 3,4,5$. 
The inner slope is set in all cases to $\alpha = 1$. 
The implied local density of dark matter is 
similar for each value of $\gamma$, and a larger scale radius is attained for steeper outer density 
slopes. This scaling of $r_0$ with $\gamma$ is ultimately a reflection of the fact that the ratio 
of the escape velocity to the LSR velocity is bound to be $2.0 \lsim v_{\text{esc}}/v_{\text{lsr}} \lsim 3.5$.

\begin{figure}[t] 
   \centering
   \includegraphics[width=3.5in]{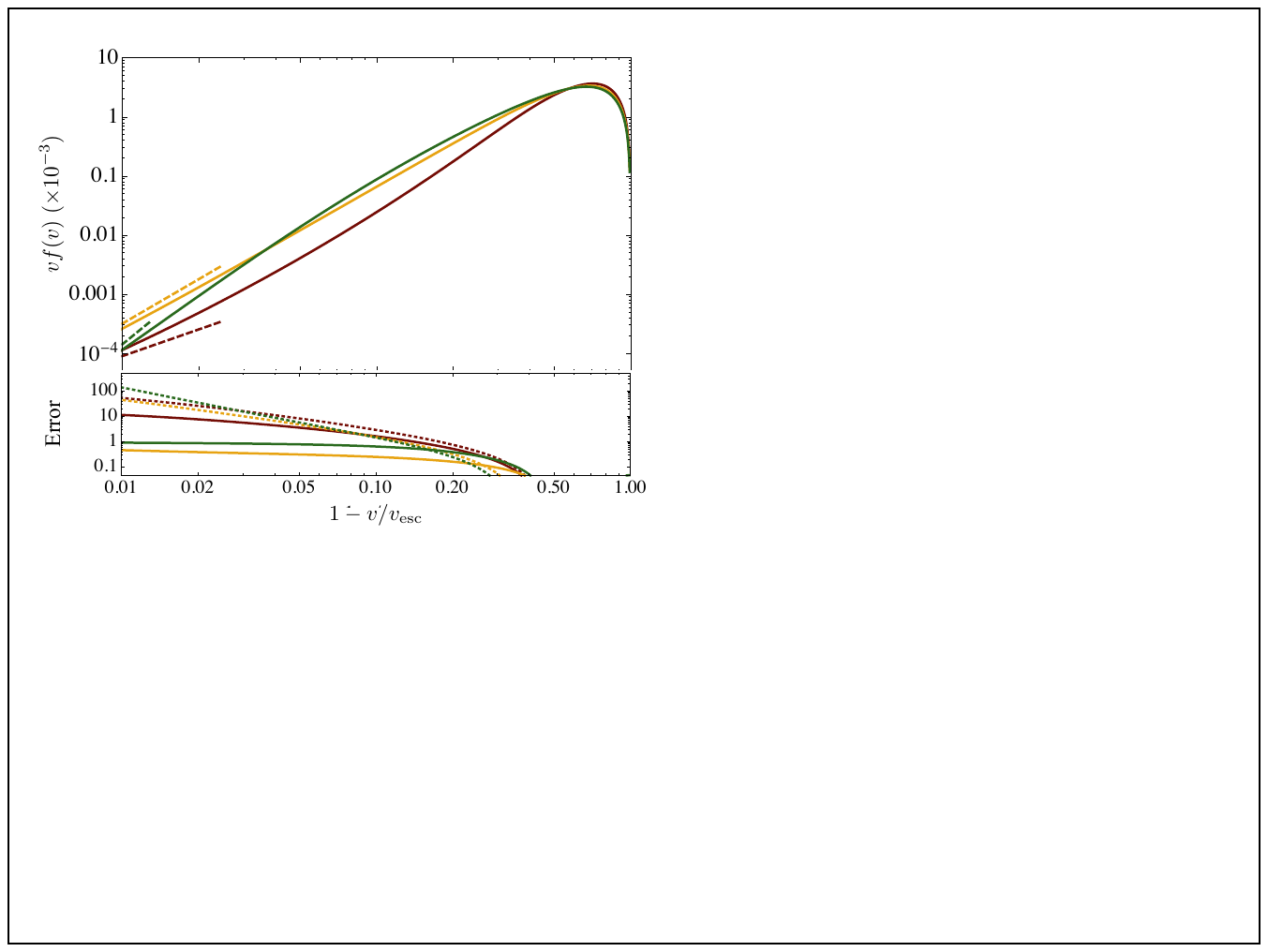} 
   \caption{Distribution functions for models with $\alpha = 1$ and $\gamma = 3$ (red), $4$ (yellow), 
   and $5$ (green) obtained by numerically solving the Eddington equation with a disk potential 
   as in (\ref{eq:disk}).  The dashed lines have slopes of $\gamma - 3/2$.  The bottom panel shows the error of using the ansatz velocity distribution (solid) or the King model (dotted) instead of the complete numerical solution.  }
   \label{fig:example}
\end{figure}

Fig.~\ref{fig:example}  shows velocity distributions, $v f(v)$, as a function of $1-v/v_{\rm esc}$ resulting from the numerical solution of the Eddington inversion formula  for the three density models with outer slopes of $\gamma = 3,4,5$ using the regions delineated in Fig.~\ref{fig:regions}.
These fits are of dark matter halos with  inner slopes of $\alpha=1$. 
The disk and bulge potential are included in the gravitational potential.  
The numerical evaluation of the Eddington formula extends over a wide enough range to ensure that the velocity distribution function is convergent; in practice, this requires that the potential goes to zero at $r \gsim 100 r_s$. 
 Tangent curves show the analytic behavior of the outer
slopes. The inclusion of the disk component does not alter
the behavior at small $1-v/v_{\rm esc}$.  The bottom panel of
Fig.~\ref{fig:example} shows the error of using the ansatz velocity
distribution function of (\ref{eq:vdist}) or the King model instead of
the full numerical solution of the Eddington equation,
$f_{\text{E}}(v)$.  The error is defined as
\begin{equation}
\text{Error} = \Bigg|\frac{ f_k(v) - f_{\text{E}}(v)}{f_{\text{E}}(v)}\Bigg|.
\end{equation}
The agreement between the power-law slope
and the numerical calculation is the best for the density models with the steepest 
outer slopes, while numerical calculation for the NFW model with outer slope behavior of $r^{-3}$ 
has a slightly steeper slope than $k = 1.5$. For the NFW, this 
is a reflection of the fact that the tail of the distribution is modified, so that as $\EE \rightarrow 0$, 
the distribution function scales as 
\begin{eqnarray}
f(\EE) \rightarrow \frac{\EE^{\frac{3}{2}}}{(-\ln \EE)^3},
\label{eq:nfweta}
\end{eqnarray}
rather than as a pure power-law in $\EE$~\cite{Widrow2000}.  The behavior
near $\EE=0$ is steeper than a $k=\frac{3}{2}$ power-law and $k_{\text{NFW}} \simeq 2.0$ typically fits better numerically.

\begin{figure*}[t] 
   \centering
   \includegraphics[width=7in]{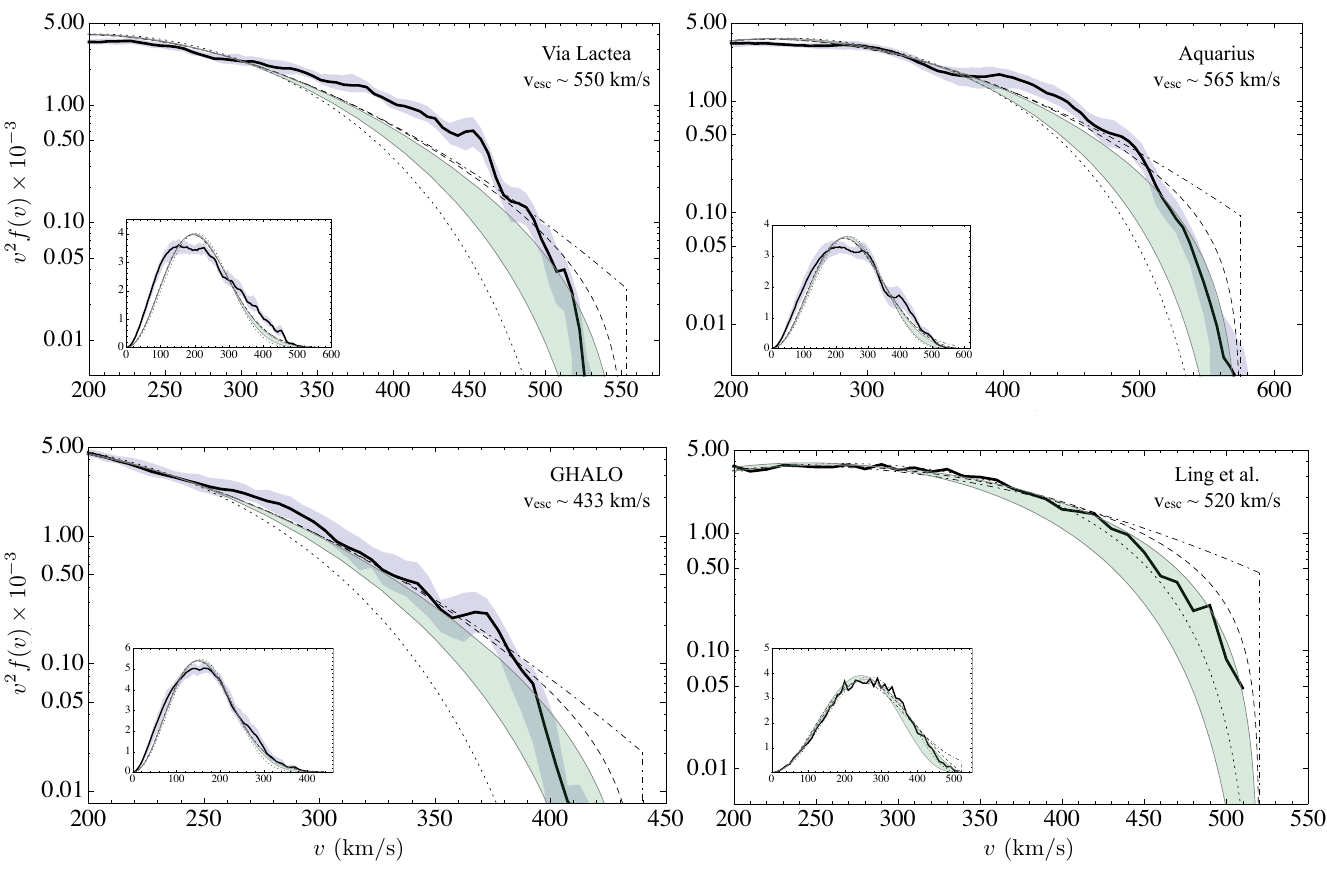} 
   \caption{Best-fit Maxwellian (dot-dashed), King (dashed), and Tsallis (dotted) 
   models to the Via Lactea, Aquarius, GHALO, and Ling et al.~\cite{Ling:2009eh}
    simulations (solid black).  The first three are simulated with
    dark matter only; \cite{Ling:2009eh} simulates a galaxy with
    baryons.  The green 
   band shows the velocity distribution obtained by fitting (\ref{eq:vdist}) to the data 
   for $1.5 < k < 3.5$.  The purple shaded regions enclose 68\% of all measured 
   distributions in each of the Aquarius, Via Lactea, and GHALO panels.
      \label{fig:vcomp}}
\end{figure*}

\subsection{Comparison to Simulated Galactic Halos} 

High resolution simulations of Galactic halos with dark matter only
have revealed a non-Maxwellian structure to velocity
distributions~\cite{Vogelsberger:2008qb,Kuhlen:2009vh}.  Several
trends are seen in these simulations.  The low velocity tail is more
populated than the best fitting Maxwellian distribution, while the
peak is more depressed.  At moderately high velocities beyond the
peak, the distribution is more populated than the best fitting
Maxwellian. Further, broad bumps that represent dynamical features are
prevalent in the velocity modulus distribution, though they are not
present in the distribution of the individual velocity components.
Although features in the distribution due to substructure are present
at some locations in the halo, their contribution to the velocity
distribution averaged over all radial shells is typically sub-dominant
relative to the broader bumps.  The highest resolution simulation in
the literature that includes baryonic physics~\cite{Ling:2009eh} does
not show such features, and is better described by a Maxwellian
distribution than the dark matter only simulations, though it still
deviates at the high velocity tail.  It remains to be seen whether
this result is robust to changes in uncertain baryonic physics and
scatter in formation histories between galaxies.

The features of these velocity distributions clearly reflect the
formation process of each individual halo, and deviate from sphericity
and isotropy in both the density distributions and velocity
ellipsoids. Of course, they also reflect the distribution of the
individual dark matter halo that was simulated, and it is not yet
known how this quantity in individual halos relates to the
distribution of the velocity distribution for a larger sample of
similar mass halos. Further, it is also assumed that
the velocity distributions from numerical simulations provide an unbiased tracer of the
velocity distribution of the actual Milky Way dark halo.  This latter
point can be addressed by choosing simulated halos to have a mass and
isolation criteria that resemble the Milky Way regime.  These
selection criteria could be further refined by demanding additional
constraints, {\em e.g.}, requiring a similar number of massive,
Magellanic cloud-like satellites to the Milky Way \cite{Busha10}.  In
spite of these caveats, comparisons to N-body simulations are still
critical for determining how well analytic arguments based upon the
assumption of equilibrium match typical galaxies.

Fig.~\ref{fig:vcomp} shows the best fits of the velocity distribution
function from \eqref{eq:vdist} to the results of the Aquarius, Via
Lactea II, and GHALO dark matter only simulations, and the simulation
with baryons of Ling et al.~\cite{Ling:2009eh}.  The high energy tails
are shown in the main figures in each of the four panels while the
insets show the fits to the full velocity distribution function.  The
shaded bands in this figure represent slopes in the range $k=[1.5,
3.5]$, corresponding to outer slopes between $\gamma= [3.0,
5.0]$.  For each value of $k$ within this band, the parameter $v_0$
has been fit to the distribution, and $v_{\rm esc}$ is set to the
value determined by each of the respective simulations. Implicit in
this assumption is that the highest velocity particle is a faithful
tracer of $v_{\rm esc}$, rather than a contaminating particle that is
not bound to the halo.

The best-fit SHM, King, and Tsallis distributions are shown for
comparison in all panels of Fig.~\ref{fig:vcomp}.  In all panels, the
SHM does not go to zero smoothly at the escape velocity, indicating
that this distribution function is unphysical in the tail.  By
construction, the velocity distribution in~\eqref{eq:vdist} smoothly
joins to a Maxwellian distribution for any value of $k$, with the
differences manifest only near the tails.

The fits shown in Fig.~\ref{fig:vcomp} indicate that the velocity
distribution in \eqref{eq:vdist} captures the high velocity tail of
the simulated distribution well, especially in comparison to the
King, SHM, and Tsallis distributions. This is in spite of the fact
that the simulations are clearly anisotropic and non-spherical, and
that the outer regimes of the halo may not have fully equilibrated.  The agreement demonstrated in
Fig.~\ref{fig:vcomp} is promising, though larger samples of dark
matter halos at similar resolution are needed to verify the high
velocity suppression that \eqref{eq:vdist} predicts.  Consistent with
previous findings, the dark matter only simulations are more
`flat-topped' than all of the models, while there are more particles
in the low regime of the velocity distribution.  The enhancement of
low velocity particles may result from radially-biassed anisotropy, as
discussed below.  The Ling et al.~\cite{Ling:2009eh} distribution best
resembles a Maxwellian profile at low velocities.  This agreement
might be attributed to either the effect of baryonic physics, which
has been shown to result in more isotropic inner halos \cite{gas}, or
to the lower resolution of dark matter simulations that include baryons.

A final point regarding the comparison to the simulations is that,
since the tail appears well described by $k = [1.5, 3.5]$, it is
plausible to conjecture that baryons do not appear to greatly affect the
shape of the velocity distribution near the escape velocity.  One way
to understand this is that since highly energetic particles with large
peri-center orbits shape the tail of the distributions, these dark
matter particles will be outside the halo's scale radius for a large
fraction of their orbits. The scale radius ranges between $r_s \simeq [10, 60] \text{ kpc}$
 (see Fig.~\ref{fig:regions}) for the models considered, and the Galactic disk is located in the regime where the density
scales as the inner slope, $\alpha$.  The inner slope should not
significantly perturb the orbits of the halo's most energetic
particles that make up the high velocity tail.  Thus, the regime most
likely to be affected by the formation of the disk are the orbits in
the the inner core of the halo, where the power-law depends on
$\alpha$.

\section{Direct Detection}
\label{sec: directdetection}

The velocity distribution function in~\eqref{eq:vdist} reproduces the results from N-body simulations and exhibits non-Gaussian behavior near the escape velocity.  This section shows that the suppression in the high energy tail has important implications for the interpretation of direct detection experiments.  
\begin{figure}[tb] 
   \centering
   \includegraphics[width=3.5in]{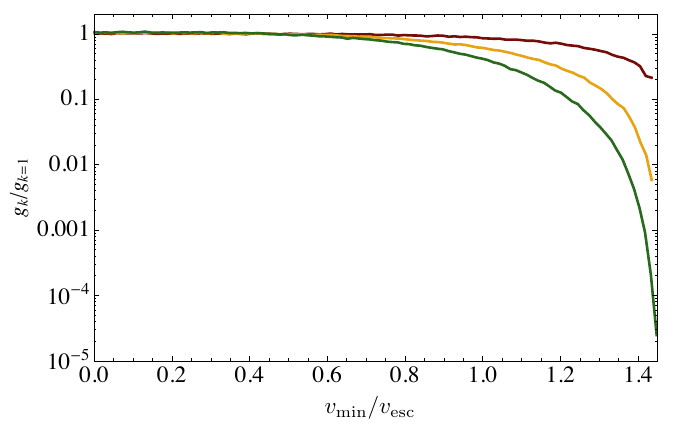} 
   \caption{Fractional change in the differential dark matter detection rate as a
     function of the minimum velocity for $k=1.5$ (red), $2.5$
     (yellow), and $3.5$ (green), compared to a King velocity
     distribution function with $k=1$.  In this figure, $v_0 = 220$
     km/s, $v_{\text{esc}} = 550$ km/s, and the Earth's velocity is
     taken at $\sim$ June 2. }
   \label{fig: vminplot}
\end{figure}

The scattering rate of a dark matter particle off a target nucleus
depends on the velocity-averaged differential cross section and is
thus sensitive to the distribution function.  The detection rate per
unit detector mass is
\begin{equation}
\frac{dR_{k}}{dE_R} = \frac{\rho_0}{m_{\text{dm}} m_N} \int_{v_{\text{min}}}^{v_{\text{esc}}}\!\!\!d^3v\; f_{k}(\vec{v} + \vec{v}_E(t)) v \frac{d\sigma}{dE_R} ,
\label{eq: diffrate}
\end{equation}
where $m_N$ is the mass of the target nucleus, $m_{\text{dm}}$ is the dark matter mass, $\mu_N$ is the reduced mass of the nucleus-dark matter system, $E_R$ is the nuclear recoil energy, $\rho_0$ is the local dark matter density, and $f_{k}(\vec{v} + \vec{v}_E(t))$ is the velocity distribution function in the Earth's rest frame with a power-law index, ${k}$ \cite{Lewin:1995rx, Schoenrich:2009bx}.  The spin-independent differential cross section is parameterized as
\begin{equation}
\frac{d\sigma}{dE_R} = \frac{m_N \sigma_n}{2 \mu_n^2 v^2} (f_p Z + f_n (A-Z))^2 |F_N(q^2)|^2,
\end{equation}
where $\sigma_n$ is the dark matter-nucleon cross section at zero momentum transfer, $\mu_n$ is the dark matter-nucleon reduced mass, and $q^2 = 2 m_NE_R$ is the momentum transfer.  $f_{p,n}$ parameterize the coupling to the proton and neutron and are typically set to unity.  $F_N(q^2)$ is the nuclear form factor, for which the Helm/Lewin-Smith form is used~\cite{Lewin:1995rx}. 
\begin{figure*}[tb] 
   \centering
   \includegraphics[width=7.0in]{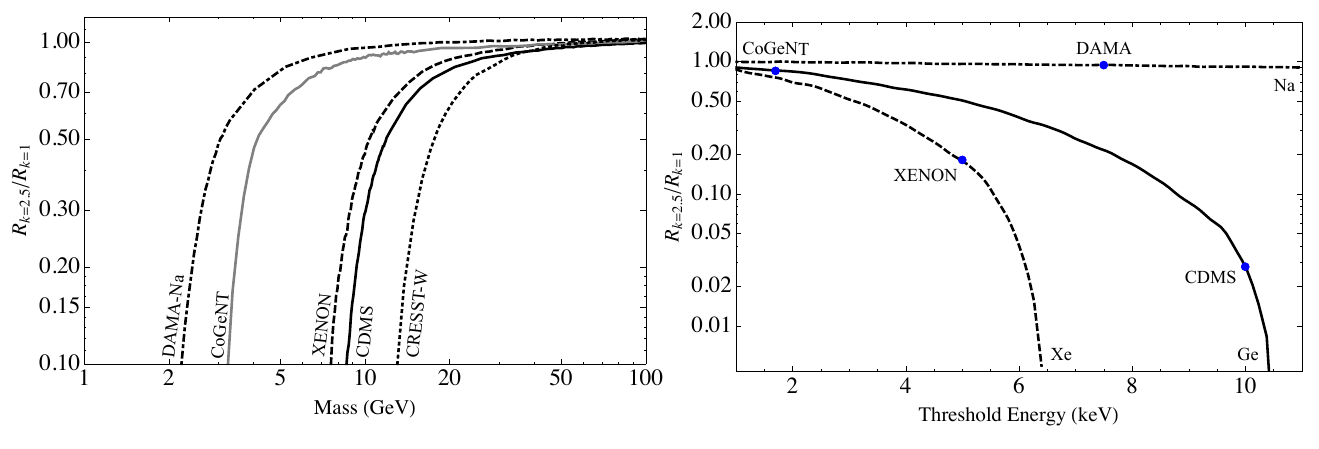}    
    \caption{Fractional change in the scattering rate for ${k} = 2.5$
      compared to ${k} =1$ for an elastically scattering dark matter
      with $v_0 = 220$ and $v_{\text{esc}} = 550$ km/s.  The plot on
      the left illustrates the dependence of the scattering rate on
      the dark matter mass for DAMA-Na (dot-dashed)~\cite{DAMA}, CoGeNT (solid
      gray)~\cite{COGENT}, XENON (dashed)~\cite{XENON100,XENON10iDM}, CDMS (solid black)~\cite{CDMS}, and CRESST-W
      (dotted)~\cite{CRESSTII} for threshold energies of 7.5, 1.7, 5, 10, and 10 keV, respectively.  The plot on the right illustrates the dependence of
      the scattering rate on the threshold energy for an 8 GeV dark matter scattering elastically off a Xe (dashed),
      Ge (solid), and Na (dot-dashed) target.  In both plots, the
      Earth's velocity was taken at $\sim$ June 2. }
   \label{fig: MEplots}
\end{figure*}

The relevant quantity when comparing direct detection rates for velocity distributions with different spectral indices is
\begin{equation}
g_k(v_{\text{min}}) = \int_{v_{\text{min}}}^{v_{\text{esc}}}\!\!\!dv\;\,  v\, f_k(\vec{v} + \vec{v}_E(t)).
\label{eq: gv}
\end{equation}
The ratio $g_k/g_{1}$ illustrates how the new {\it ansatz} compares with the more commonly used King model.  In the limit where $v_{\text{min}} \rightarrow 0$, the ratio $g_k/g_{1}$ approaches unity for properly normalized distributions.  However, this ratio deviates from unity when the minimum scattering velocity is $\sim 60\%$ of the escape velocity, as illustrated in Fig.~\ref{fig: vminplot}.
As the minimum velocity approaches the maximum dark matter velocity, $v_{\text{esc}} + v_E(t)$, the scattering rate is strongly suppressed.  This suppression can be as large as two orders of magnitude for an outer slope of $\gamma= 4$, or five orders of magnitude for $\gamma = 5$.

Dark matter with particularly large minimum scattering velocities will be most sensitive to increases in the spectral index of the velocity distribution function.  
The minimum velocity depends on the kinematics of the scattering event and is given by
\begin{equation}
v_{\text{min}} = \frac{1}{\sqrt{2 m_N E_R}} \Big( \frac{m_N E_R}{\mu_N} + \delta\Big),
\label{eq: vmin}
\end{equation}
where $\delta$ is the mass splitting between the initial and final dark matter states if dark matter scattering is dominantly inelastic; elastic scattering corresponds to $\delta = 0$. 
Figure~\ref{fig: vminplot} shows the effect of the power-law index $k$ on the direct detection rate, as a function of $v_{\text{min}}$.

Inelastic dark matter~\cite{inelasticDM} is sensitive to the spectral index of the distribution function because it has a larger $v_{\text{min}}$.  The minimum scattering velocity  differs between experiments with different energy thresholds or target nuclei.  As a result, a change in the spectral index of the velocity distribution affects the limits for each experiment differently.   For example, an experiment with a heavier target nucleus or a lower threshold energy will have a fairly large value for $v_{\text{min}}$ and a greater sensitivity to the velocity distribution function. 

Light elastic dark matter is also sensitive to the high velocity tail
of Galactic dark matter~\cite{lightDM}.  Fig.~\ref{fig: MEplots}
illustrates the suppression of the scattering rate $R_k$ (integrated
over recoil energy) relative to that for the King model.  For example,
if $m_{\text{dm}}=8 \GeV$, a Xe target with 5~keV threshold energy has
$R_{2.5}/R_1 \simeq 0.18$.  This ratio is more suppressed as one goes
to heavier target nuclei and/or larger threshold energy.  For
instance, a Ge target with $E_{\text{th}}= 10 \keV$ is suppressed
relative a Ge target with a $E_{\text{th}}=2 \keV$ by a factor of 25.
In the case of light target nuclei (i.e., Na), or low threshold
energies (i.e., $\lsim 2\keV$), the effect of increasing the spectral
index is much less significant, with $R_{2.5}/ R_1~\sim 1$.  The
relative changes in the rates have profound consequences when
comparing a potential signal in one experiment with the result of
another experiment with a different target nucleus or threshold
energy.
 
\section{Discussion and Conclusion} 
\label{sec:outlook}
This paper has considered a cosmologically-motivated class of double
power-law dark matter density profiles and has examined the resulting
velocity distribution, assuming isotropy and spherical symmetry. This
article has specifically focused on the behavior of the velocity
distribution at low binding energies, near the escape velocity. The
primary result is that the velocity distribution function is
well-modeled by the function presented in \eqref{eq:vdist}, where the
power-law index ,$k$, is in the range $\sim 2$. This is a steeper
fall-off than what is expected for a Maxwell-Boltzmann distribution,
which has served as the canonical model for dark matter velocity
distributions in the recent literature.  The new ansatz in
\eqref{eq:vdist} affects the interpretation of limits set by direct
dark matter detection experiments. Limits on low mass dark matter,
$m_{\text{dm}}\lesssim 40\GeV$, are most strongly affected, especially
for experiments with heavy target nuclei or large threshold energies.
 
The analysis presented in this article does not address the effects of
anisotropy and deviations from spherical symmetry, which 
are expected to be more serious in the outskirts of the profile
most relevant for the high velocity tail~\cite{Allgood:2005eu}.  For instance,
anisotropy could be responsible for $\alpha \ll 2$.  For anisotropic
distributions, there is no guarantee that a given density distribution
will correspond to a unique distribution function.  For example, if
the distribution function is deformed to $L^{-2\beta}f(\EE)$, the
distribution function depends only on a single derivative of the
density with respect to the potential for radial orbits of $\beta =
\half$ \cite{Evans:2005tn}.  The corresponding velocity distribution
scales as $f(v) \propto v$ at low velocities, which is broadly
consistent with the behavior in the dark matter only simulations of
Fig.~\ref{fig:vcomp}.  In comparison, the low velocity behavior of the
Galactic halo for a circularly biased system with $\beta = -\half$ is
$f(v) \propto v^3$.  Thus, the low energy tail may indicate the
presence of radial orbits in the halos, which would be consistent with
the velocity anisotropy distribution in numerical
simulations~\cite{Hansen:2004qs}.

The interplay between density profiles of dark matter halos and their
corresponding velocity distribution functions was developed in this
paper, but requires further study.  Exploring equilibrium dark matter
phase space distributions, particularly anisotropic distributions, may
give further insight into the behavior of N-body simulations and may
explain the inner slope of $\rho(r)$.  These effects are unlikely to
be directly observable in early direct detection experiments, but may
be seen in future directional dark matter detection
experiments~\cite{Host:2007fq}.

\section*{Acknowledgements}
We thank Sonia El-Hedri and Daniele Alves for enlightening discussions
on the behavior of Maxwell-Boltzmann-like distributions at large and
small radii.  We thank Patrick Fox, Jeremiah Ostriker, David Spergel,
and Neal Weiner for useful discussions. We additionally thank Michael
Kuhlen, Fu-Sin Ling, and Mark Vogelsberger for providing numerical
files used in Fig.~\ref{fig:vcomp}.  JGW and RHW are supported by the
DOE under contract DE-AC03-76SF00515.  JGW is partially supported by
the DOE's Outstanding Junior Investigator Award and the Sloan
Foundation.  LES acknowledges support for this work from NASA through
Hubble Fellowship grant HF-01225.01 awarded by the Space Telescope
Science Institute, which is operated by the Association of
Universities for Research in Astronomy, Inc., for NASA, under contract
NAS 5-26555.  ML acknowledges support from the Simons Postdoctoral
Fellows Program and the LHC Theory Initiative.  Lastly, we acknowledge the second sphenic number for inspiration.

\end{document}